\documentclass[12pt]{JHEP3}

\usepackage[numbers,square,comma,compress]{natbib}

\usepackage{epsfig,multicol,bbm}

\newcommand\fverbdo{\egroup\medskip\noindent%
            \fbox{\unhbox\fverbbox}\ }
\newcommand\fverbit{\egroup\item[\fbox{\unhbox\fverbbox}]}
\newbox\fverbbox

\title{\bf  Counting to one:
reducibility of one- and two-loop amplitudes at the integrand level}

\author{Ronald H.P. Kleiss$^{a}$\footnote{R.Kleiss@science.ru.nl}\,, 
Ioannis Malamos$^{b}$\footnote{Supported by REA Grant Agreement PITN-GA-2010-264564 (LHCPhenoNet),
by the MICINN Grants FPA2007-60323, FPA2011-23778 and
Consolider-Ingenio 2010 Programme CSD2007-00042 (CPAN)}\footnote{Ioannis.Malamos@ific.uv.es}\,,
Costas G. Papadopoulos$^{c,d}$\footnote{costas.papadopoulos@cern.ch}\,,
Rob Verheyen$^{a}$\footnote{robverheyen@gmail.com}

\\

$^a$~{Radboud University, Nijmegen, The Netherlands}\\
$^b$~{Instituto de F\'{\i}sica Corpuscular, \\ Consejo Superior de Investigaciones Cient\'{\i}\-fi\-cas-Universitat de Val\`encia, \\
Parc Cient\'{\i}fic, E-46980 Paterna (Valencia), Spain.}\\
$^c$~{NCSR `Demokritos', Agia Paraskevi, 15310, Greece}\\
$^d$~{Department of Physics, Theory Division, CERN, 1211 Geneva 23, Switzerland}\\

}

\abstract{  Calculation of amplitudes in perturbative quantum field theory involve large loop integrals.
The complexity of those integrals, in combination with the large number of Feynman diagrams, 
make the calculations very difficult. Reduction methods proved to be very helpful, lowering 
the number of integrals that need to  be actually calculated. Especially, the reduction at the 
integrand level technique, 
improves the speed and set-up of these calculations.
In this article we demonstrate, by counting the numbers of tensor structures and independent
coefficients, how to write such relations at the integrand level for one$-$ and two$-$loop
amplitudes. We clarify their connection to the so-called spurious terms at one loop and discuss their structure
in the two$-$loop case. This  method is also applicable to higher loops, 
and the results obtained apply to both planar and non-planar diagrams.}

\preprint{
          }

\begin{document}

\newcounter{im}
\setcounter{im}{0}
\newcommand{\exampleSp}{\stepcounter{im}\includegraphics[scale=0.9]{SpinorExamples_\arabic{im}.eps}}
\newcommand{\myindex}[1]{\label{com:#1}\index{{\tt #1} & pageref{com:#1}}}
\renewcommand{\topfraction}{1.0}
\renewcommand{\bottomfraction}{1.0}
\renewcommand{\textfraction}{0.0}
\newcommand{\nn}{\nonumber \\}
\newcommand{\eqn}[1]{Eq.~\ref{eq:#1}}
\newcommand{\be}{\begin{equation}}
\newcommand{\ee}{\end{equation}}
\newcommand{\ba}{\begin{array}}
\newcommand{\ea}{\end{array}}
\newcommand{\bea}{\begin{eqnarray}}
\newcommand{\eea}{\end{eqnarray}}
\newcommand{\bino}[2]{\left(\begin{tabular}{c}$#1$\\$#2$\end{tabular}\right)}
\newcommand{\bqa}{\begin{eqnarray}}
\newcommand{\eqa}{\end{eqnarray}}
\newcommand{\nl}{\nonumber \\}
\def\db#1{\bar D_{#1}}
\def\zb#1{\bar Z_{#1}}
\def\d#1{D_{#1}}
\def\tld#1{\tilde {#1}}
\def\slh#1{\rlap / {#1}}
\def\eqn#1{eq.~(\ref{#1})}
\def\eqns#1#2{Eqs.~(\ref{#1}) and~(\ref{#2})}
\def\eqnss#1#2{Eqs.~(\ref{#1})-(\ref{#2})}
\def\fig#1{Fig.~{\ref{#1}}}
\def\figs#1#2{Figs.~\ref{#1} and~\ref{#2}}
\def\sec#1{Section~{\ref{#1}}}
\def\app#1{Appendix~\ref{#1}}
\def\tab#1{Table~\ref{#1}}
\def\cg{c_\Gamma}
\newcommand{\bfig}{\begin{center}\begin{picture}}
\newcommand{\efig}[1]{\end{picture}\\{\small #1}\end{center}}
\newcommand{\flin}[2]{\ArrowLine(#1)(#2)}
\newcommand{\ghlin}[2]{\DashArrowLine(#1)(#2){5}}
\newcommand{\wlin}[2]{\DashLine(#1)(#2){2.5}}
\newcommand{\zlin}[2]{\DashLine(#1)(#2){5}}
\newcommand{\glin}[3]{\Photon(#1)(#2){2}{#3}}
\newcommand{\gluon}[3]{\Gluon(#1)(#2){5}{#3}}
\newcommand{\lin}[2]{\Line(#1)(#2)}
\newcommand{\sof}{\SetOffset}

\section{Introduction}

Modern colliders such as the Large Hadron Collider (LHC) (and the Tevatron before it was shut down)
produce a large amount of experimental data.
In order to understand the output of these experiments, comparison between 
very precise theoretical results and experimental results is needed. 
It is clear, from the theoretical point of view, that Next-to-Leading-Order 
(NLO) and  Next-to-Next-to-Leading-Order (NNLO)
calculations with many external legs have to be considered~\cite{Anastasiou:2009zz}.
This implies that (very) large loop integrals have to be computed for very
many Feynman diagrams, which has widely been considered the bottleneck of
such calculations.

Reduction techniques form a way out. The idea of reducing Feynman integrals with a
large number of denominators to a set of simpler integrals 
({\it i.e.\/} with fewer denominators) at one loop goes surprisingly many 
years back~\cite{Kallen:1964zz,Melrose:1965kb}.
A typical integral with $n$ such denominators, in $d$ space-time dimensions, is given by
\[
\int d^d q \frac{1}{D_1D_2...D_n}\;\;,
\]
where $D_i=(q+p_i)^2-m_i^2$ is the denominator of a generic propagator.
In \cite{Kallen:1964zz} the 
authors reduce a triangle (integral with 3 denominators) to bubbles (2 denominators)
in 2 dimensions while in \cite{Melrose:1965kb} a pentagon (5 denominators) is 
reduced to boxes (4 denominators) in 4 dimensions. We see that the result of the reduction 
depends on the dimension. However, the methods we will use can 
be applied to all dimensions. Our main interest is of course the case $d=4$.

Since a few decades now~\cite{'tHooft:1978xw}, it is a very well known fact that a generic one-loop
amplitude is decomposable in terms of scalar integrals, with one, two, three and four external legs (in $d=4$).
Passarino and Veltman \cite{Passarino:1978jh} used  Lorentz invariance to express tensor one-loop $n-$point integrals 
in terms of $m-$point scalar integrals ($m\le n$). As a consequence, 
only the evaluation of scalar integrals (integrals with trivial numerators) 
is needed in order to perform a one-loop calculation. 

In another attempt~\cite{vanNeerven:1983vr} a pentagon-to-boxes decomposition is performed in 4 dimensions. 
The importance of this paper is that it provides a basis in four-dimensional momentum space
(the so called van Neerven-Vermaseren basis), which proved useful for understanding 
one-loop reduction. Another important fact about 
this paper is the use of what we call nowadays {\em spurious terms\/}
to decompose a scalar 
pentagon to boxes. Spurious terms are  terms that, by construction, vanish upon integration. Their r\^ole
will be explained later when we consider reductions at the integrand level and we
shall see why one cannot avoid them. 

The next big step comes from unitarity methods \cite{Bern:1994zx,Bern:1994cg,Bern:2005hs,Bern:2005ji,Bern:2005cq}. 
Instead of working with specific Feynman diagrams these methods have a big advantage 
in that they try to decompose the whole one-loop amplitude in terms of the scalar integrals. By cutting 
propagators\footnote{`Cutting' a propagator means that loop momenta are
chosen for which one or more of the $D$'s vanish so that the integrand
becomes singular. One also speaks of `putting propagators on-shell'.} 
the rational coefficients of loop integrals are given 
in terms of products of tree amplitudes. In generalized unitarity methods~\cite{Britto:2004nc,Britto:2005fq,Britto:2004ap,Anastasiou:2006jv,Berger:2006cz,Giele:2008ve}, 
the notion of {\em multiple cuts\/} is introduced. 
One can cut more than one propagator to find these coefficients. Note that,
for $d=4$, cutting four propagators essentially determines the
loop momentum (there is, in general, more than a single solution since the
$D$'s are quadratic in the loop momentum).

The Ossola-Papadopoulos-Pittau (OPP) method \cite{Ossola:2006us,Ossola:2007bb,Ossola:2007ax,Ossola:2008xq,Mastrolia:2008jb} comes as 
a natural combination of all the above. Since every integral can be 
decomposed to scalar integrals with up to four denominators (for $d=4$), 
every one-loop amplitude is written in terms of coefficients that multiply these scalar integrals. 
The OPP method works at the integrand level~\cite{Forde:2007mi,Ellis:2011cr}, which means that for these decompositions to 
be possible one must also include  spurious terms.
Then one has to find a way to calculate the coefficients of the
reduction and  multiply them with the appropriate scalar integrals, using one of 
the packages available for the evaluation 
of them(i.e \cite{vanHameren:2010cp,Ellis:2007qk}). 
Finding the coefficients is a purely algebraic problem. 
The method is suitable for a fully numerical implementation. 
The OPP method has  been widely used so far in many one loop calculations (see
for example \cite{Binoth:2008kt,vanHameren:2009dr,Bevilacqua:2009zn,Bevilacqua:2010ve,Bevilacqua:2010qb,Kardos:2011qa,Bevilacqua:2011aa,Kardos:2011na,Garzelli:2011is,
Hirschi:2011pa,Frederix:2011qg,Frederix:2011ss,Frederix:2011ig}).

As we noticed before, in the case of one-loop calculations a basis for any integral is known in advance. Any
one-loop integral can be written in terms of scalar
boxes, triangles, bubbles and tadpoles. However,
in the case of higher-loop integrals the situation is different. 
A basis is not known {\it a priori}. It is believed that unitarity methods can also be applied in that case 
and there are some recent papers in 
that direction, performing decomposition \`a la OPP~\cite{Mastrolia:2011pr,Badger:2012dp,Zhang:2012ce,Mastrolia:2012an},
or using generalized unitarity~\cite{Gluza:2010ws,Kosower:2011ty,Schabinger:2011dz} at two loops. 

Two remarks are in order here. The first is that
the basis of two-loop
integrals does not include only scalar integrals. It includes integrals that also have
irreducible scalar products (ISP) as numerators (to some power) that cannot be rewritten as 
existing denominators of the integral. In the one-loop case these ISP 
are always spurious and integrate to zero, but for higher loops this does no longer hold. The second remark is that
if one is interested in constructing a unitarity-like basis, the set of integrals that ends up with
is not necessarily a minimal one: the integrals are not by default Master Integrals (MI). 
There might be 
smaller sets of true MI and at two or more loops one can find them by using integration-by-parts (IBP)
identities \cite{Tkachov:1981wb,Chetyrkin:1981qh,Smirnov:2004ym}. 

In this article we  prove that any 
two-loop integral can be written as a linear combination of integrals with 
at most $2d$ denominators. From this set of integrals, one can in principle end 
up with true master integrals with numerators containing ISP 's.
The layout of this paper is the following.
We start with definitions and reduction at one loop. We see why one can have unitarity-like bases
at the integrand level by writing first the numerator of our integrals (for scalar cases the number
one) as a sum of coefficients times denominators. We investigate, using simple algebra, what 
is the form of these coefficients in order for our polynomial equation to have solutions. 
Then we repeat
the same procedure in the case of two-loop integrands. Our method is not exactly the OPP one, but the 
connection of the two methods will become clear.

\section{{Reduction at one loop}}
\subsection{Introduction : reduction with trivial coefficients}

We start with  integrands of any given one-loop amplitude. 
These integrands consist of the sum of  integrands coming from the Feynman diagrams that 
contribute to a given  process and share the same topology;
the advantage is that we perform the decomposition once instead of reducing every 
single diagram separately. 
For that reason we deal with 
integrand-graphs, or iGraphs instead of Feynman diagrams. 
We give an example of an iGraph of order 5 (pentagon) below, where $j=1\ldots 5$,
\bea
D_j \equiv D(q+p_j) = (q+p_j)^2-{m_j}^2 = q^2+2(p_j\cdot q) + \mu_j\;\;\;,
\;\;\;\mu_j\equiv {p_j}^2-{m_j}^2\;\;,
\eea

\begin{center}
\[ \epsfig{figure=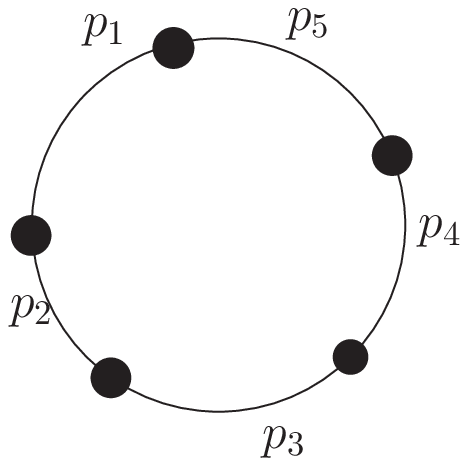,,width=5cm} \]
\end{center}
\begin{center}
{\small A one-loop pentagon iGraph. 
The dots serve to
distinguish the\\ several denominators. The external momenta are indicated.}
\end{center}

The loop momentum is denoted by $q^\mu$, 
and ${p_j}^\mu$ is called the {\em external momentum}, where it must be realized that by this we do {\em not\/} mean a momentum related to a particle incoming
or outgoing in a given amplitude: what we call external momenta are simply
fixed momenta, given in some way by the configuration of incoming and outgoing
momenta and the various diagram topologies.

Consider a one-loop iGraph of order $n$:
\[ {1\over D_1D_2D_3D_4...D_n}\;\;.\]
We say that we can decompose this iGraph
 {\em if\/} we can find  funtions $T_{1,2,...,n}(q)$ such that
\be
D_1T_1(q) + D_2T_2(q) + \cdots + D_nT_n(q) = 1\;\;,
\label{counttoone}
\ee
for then we have 
\be
{1\over D_1\cdots D_n} = {T_1(q)\over D_2D_3D_4...D_n}
+ {T_2(q)\over D_1D_3D_4...D_n} + \cdots + {T_n(q)\over D_1D_2D_3...D_{n-1}}\;\;,
\ee
and the original iGraph is decomposed into 
a sum of iGraphs of order $n-1$ (or lower). 

This immediately leads us to state the following theorem: {\em one-loop iGraphs of order
$d$ or smaller {\bf cannot\/} be decomposed in the above manner}.
The reason
is simple: for $n\le d$ there exist a cut through {\em all\/} propagators, so that
$D_j=0$ for $j=1,\ldots,n$ and \eqn{counttoone}
 then would become $0=1$. 
 
The simplest possibility for the functions
 $T_j(q)$ is to take them to be just numbers independent of $q^\mu$ 
 (`trivial' coefficients):
\[
T_j(q) = x_j\;\;.
\] 
From  \eqn{counttoone} then we have
\be
q^2 \sum_{j=1}^n x_j + 2q_\mu\sum_{j=1}^n x_j{p_j}^\mu + 
\sum_{j=1}^n x_j\mu_j = 1\;\;.
\ee
Since this has to hold for {\em any\/} value of $q^\mu$ we must have
separately
\be
\sum_{j=1}^n x_j = 0\;\;\;,\;\;\;\sum_{j=1}^n x_j{p_j}^\mu = 0\;\;,
\label{homogeneouseq}
\ee
and 
\be
\sum_{j=1}^n x_j\mu_j = 1\;\;.
\label{inhomogeneouseq}
\ee
Note that if a nontrivial solution to the homogeneous equations
(\ref{homogeneouseq}) exists, then by suitable scaling we can always
satisfy \eqn{inhomogeneouseq}. We see that, at one loop, for $d=4$ any iGraph of
order 6 or higher can be decomposed in this fairly trivial way.
A pentagon in 4-dimensions thus cannot
be decomposed that way. For general $d$,
iGraphs of order $d+2$ or higher are decomposable.

\subsection{Reduction with coefficients linear in the loop momentum}
For a one-loop iGraph of order 5 (or lower) no trivial decomposition exists
in $d=4$, assuming that four out of the five external momenta, $p_j$, in the loop can be considered as linearly independent.
One way to see this is by shifting the loop momentum so that
$\sum_j{p_j}^\mu = 0$. Then the only solution to the  conditions in
\eqn{homogeneouseq} is $x_j=0,j=1\ldots,5$ which is unacceptable in
\eqn{inhomogeneouseq}. We therefore turn to the next simplest possibility for the
$T$'s, with a linear $q$ dependence :
\be
T_j(q) = x_j + \sum_{k=1}^4 x_{j,k}(q\cdot t_k)\;\;.
\label{linearterms}
\ee
The single $x_j$ is now replaced by 5 (or $d+1$) variables to be
determined in each $T$.
Here, the four (or $d$) vectors ${t_k}^\mu$ must be  linearly independent but are 
otherwise arbitrary. In analogy to \eqn{homogeneouseq} and
\eqn{inhomogeneouseq}, we now have more tensor structures in terms of the
loop momentum : we can denote them by the shorthand
\be
1\;\;,\;\;q^\mu\;\;,\;\;q^\mu q^\nu\;\;,\;\;q^2q^\mu\;\;.
\ee
There are, for $d=4$, therefore 1+4+10+4 = 19 independent tensor
structures. Note that the $q^2$ appearing in \eqn{homogeneouseq} is no longer
independent since it appears as the trace part of $q^\mu q^\nu$.  This
can be extended to the inclusion of higher-rank tensors and other dimensions : 
in $d$ dimensions, and with the inclusion of tensor up to rank $k$, we find
for the number $N(d,k)$ of independent tensor structures
\be
N(d,k) = \bino{d-1+k}{k} + \sum_{p=0}^{k+1} \bino{d-1+p}{p}\;\;.
\ee
In the table below we give the results for various ranks and dimensionalities.
\begin{center}
\begin{tabular}{|r|ccccc|}\hline\hline
$k$      & 0 & 1 & 2 & 3 & 4 \\ \hline
$d=$1 &  3 & 4 & 5 & 6& 7 \\
        2 & 4 & 8 & 13 & 19 & 26 \\
        3 &  5 & 13 & 26 & 45 & 71 \\
        4 &  6 & 19 & 45 & 90 & 161 \\
        5 &  7 & 26 & 71 & 161 & 322 \\
        6 &  8 & 34 & 105 & 266 & 588 \\
        \hline\hline
\end{tabular}\\ {\small Values of $N(d,k)$}
\end{center}
The number of coefficients $x$ to be solved for is given by
\be
X(n,d,k) = n\bino{d+k}{k}\;\;.
\ee
Since for $d=4$ and $k=1$ we have $N(4,1)=19$, it would seem that
iGraphs of order 5 and 4 are decomposable with linear terms. However,
the situation is not so simple since it is not obvious that the 25 coefficients for
$n=5$ and the 20 coefficients for $n=4$ allow us to actually build up the 19
required tensor structures. We now describe how we can ascertain the
number of independent structures numerically, by an approach that
may be dubbed {\em cancellation probing}.

We start by generating {\em random\/} values for the external momenta
${p_j}^\mu$ and $m_j$ ($j=1,\ldots,n$). This avoids any possibility of us choosing,
coincidentally, any special phase space point where degeneracy might occur.
Then, we choose {\em random\/} values for $q^\mu$ precisely $\xi=X(n,d,k)$
times, and insert all this in \eqn{counttoone}. We are left with a set of $\xi$ linear
equations for the $\xi$ unknowns $x$ :
\be
\sum_{j=1}^{\xi} {M^i}_jx^j = 1\;\;\;,\;\;\;j=1,\ldots,\xi\;\;.
\label{systemofequations}
\ee
The $\xi\times\xi$ matrix $M$ is purely numerical. We obtain it using the
computer-algebra package {\tt MAPLE}\footnote{http://www.maplesoft.com/} which, although not
numerically the fastest available, has the essential advantage that one can
easily set the precision with which numerical operations are 
performed\footnote{The relevant
variable is {\tt Digits}.}. Now, if the number of independent tensor structures that
can be formed with our $T$'s is less than $\xi$, the determinant of $M$ will
vanish. In an ideal real-number model of computation, we would thus find
$det(M)=0$, but in our actual numerical computation there will be rounding
errors. A cancellation of numbers to `zero' will , in {\tt MAPLE}, actually give
a number of order $10^{-p}$, where $p$ is the number of digits specifies in
the precision we tell {\tt MAPLE} to use. If the matrix' determinant is computed by
Gaussian elimination\footnote{This is almost unavoidable since the matrix
$M$ is {\em not\/} sparse, and anyway we can choose Gaussian elimination
as an option in any case.}, then a matrix with $q$ zero eigenvalues will 
have a determinant of order $10^{-pq}$. By letting $p$ run down from
150 to 20 in steps of 10, we can obtain\footnote{Surprisingly, the cancellation
probing appears to fail for $p=15$ and $p=10$, possibly since {\tt MAPLE}
may have special ways to treat these accuracies ($p=10$ is default
in {\tt MAPLE}).}  a very accurate estimate of $q$, especially since $q$ must be
integer. We give two examples to demonstrate the use of this method for the
calculation of the zero eigenvalues. In both examples we consider a decomposition
of pentagon to boxes, with linear and quartic terms respectively. 
\begin{center}\begin{minipage}{6cm}{
\epsfig{figure=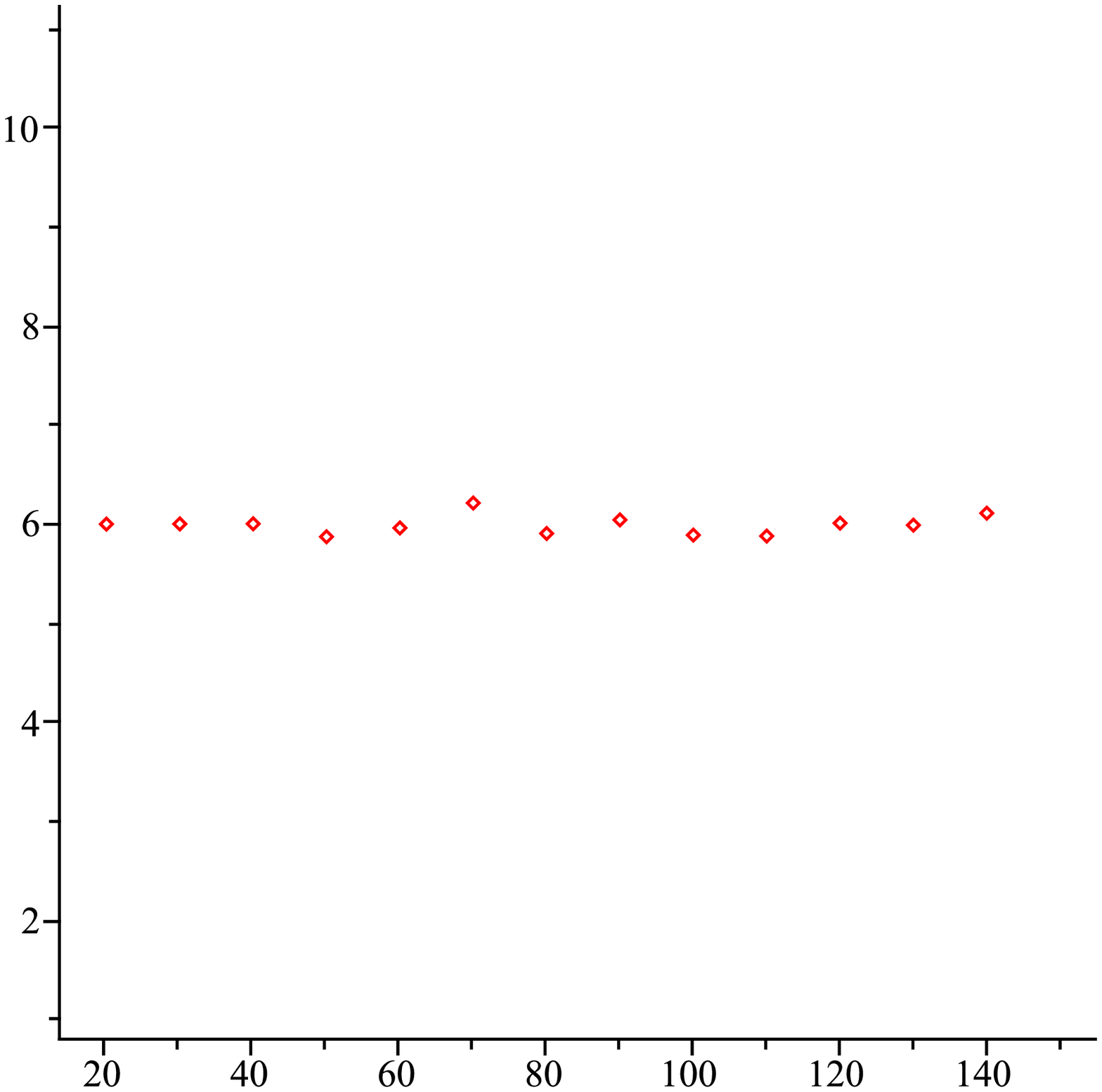,width=6cm}}
\end{minipage}
\begin{minipage}{6cm}{The number of zero eigenvalues using rounding errors
for the case of a pentagon decomposition with coefficients linear in the
 loop momentum. The obtained value is
$q = 5.988\;\pm\; 0.098$; the (absolute values of the $25\times25$
determinants range from
$1.6\;10^{-826}$ to $2.3\;10^{-108}$.}
\end{minipage}
\end{center}
\begin{center}
\begin{minipage}{6cm}{
\epsfig{figure=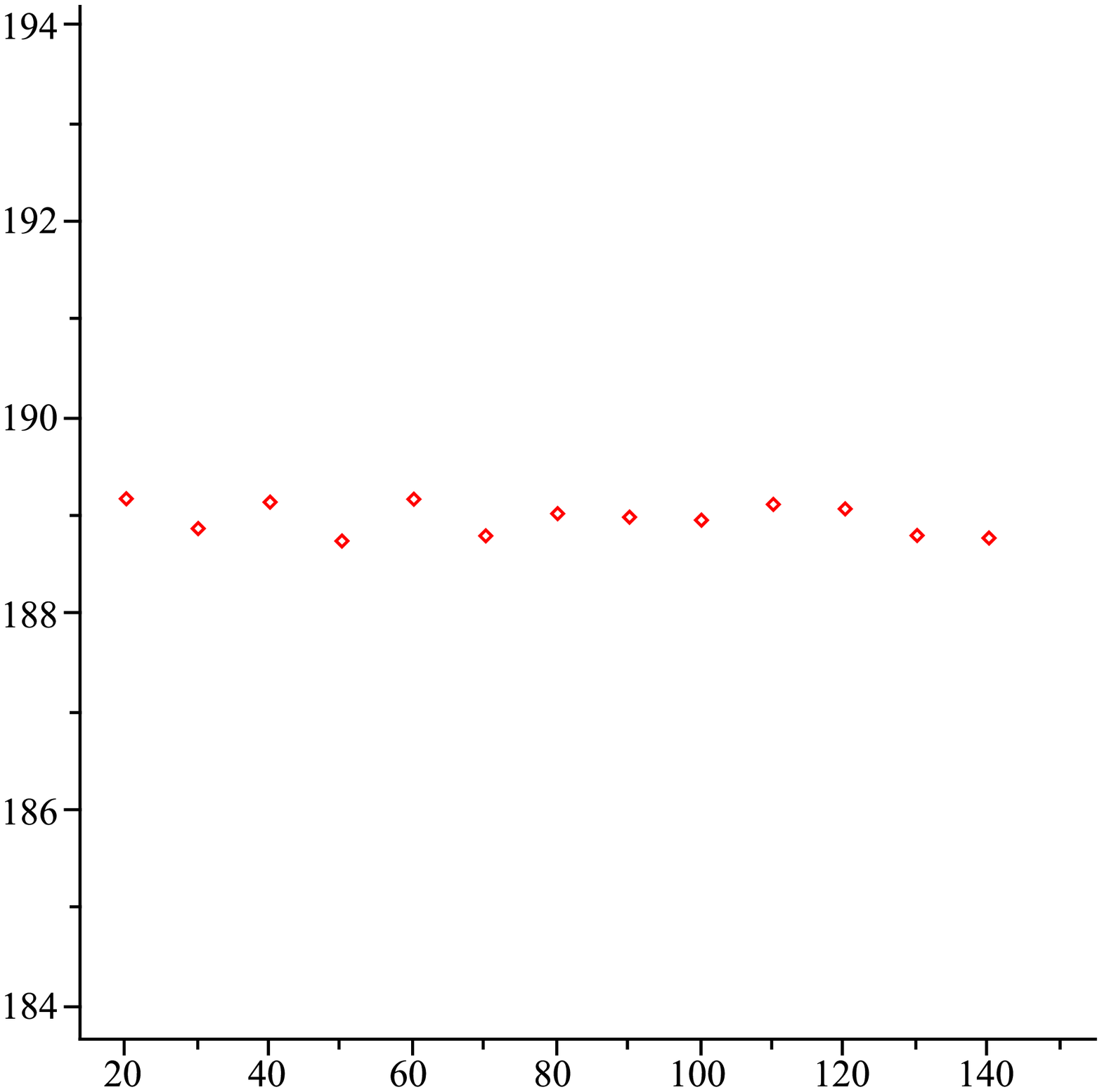,width=6cm}}
\end{minipage}
\begin{minipage}{6cm}{The number of zero eigenvalues using rounding errors for the case of
a pentagon decomposition with up to quartic in the loop momentum coefficients. 
Here, $q =  188.98\;\pm\;0.15$, with the $350\times350$ determinants ranging 
from $2.7\;10^{-26000}$ to $2.3\;10^{-3320}$.}
\end{minipage}
\end{center}
The difference $\xi-q$ then gives the rank of $M$, and {\em this\/}
determines the decomposability of the iGraph :
the rank must be equal to $N(d,k)$ for it to be
decomposable. In the table below we give
the results of cancellation probing for various $n$ and $d$.
\begin{center}
\[\begin{tabular}{|c|c|c|c|c|c|c|}
\hline\hline
$n$ & $d=6$ & $d =5$ & $d=4$ & $d=3$ & $d=2$ & $d=1$ \\ \hline
2   &  14-0  &  12-0  &  10-0  &  8-0  & 6-0     & 4-0   \\ \cline{6-6}
3   & 21-1  & 18-1  & 15-1  & 12-1  & 9-1   & 6-2   \\ \cline{5-5}
4   & 28 -3 & 24-3  & 20-3  & 16-3  & 12-4  & 8-4 \\ \cline{4-4}
5   & 35-6  & 30-6  & 25-6  & 20-7  & 15-7  &  10-6 \\ \cline{3-3}
6   & 42-10 & 36-10 & 30-11 & 24-11 & 18-10 & 12-8\\ \cline{2-2}
7   & 49-15 & 42-16 & 35-16 & 28-15 & 21-13 & 14-10\\
8   & 56-22 & 48-22 & 40-21 & 32-19 & 24-16 & 16-12\\ 
\hline\hline
\end{tabular}\]
{\small The rank of $M$  for various $d$ and $n$,\\
given as the difference $\xi-q$.}
\end{center}
We have denoted the limit of decomposability with horizontal lines. We conclude
that in four dimensions, $n=5$ is precisely decomposable, but $n=4$ is not.
We now also see the deeper reason for this:
in spite of there being 20 coefficients (one more than the minimum of 19),
only 17 independent combinations can actually be formed. We also see that
for sufficiently large $n$ the number of independent combinations of coefficients
saturates at $N(d,1)$ as it ought to. We conclude that in $d$ dimensions,
an iGraph of order $d+1$ is precisely decomposable with linear terms, but one
of order $d$ is of course not. 

In the OPP method \cite{Ossola:2006us}, the linear terms are precisely
the spurious terms. We have to note that our general linear terms are not exactly those. The spurious terms have a specific property leading to fewer tensor structures, and we give an example in the appendix.
Rewriting our general linear terms in terms of propagators and spurious terms, we see that we decompose a pentagon into boxes {\em and\/} triangles
(like in \cite{vanNeerven:1983vr}and \cite{Ossola:2006us} for example).
It can be checked that the triangles  always cancel, and therefore the decomposition
is actually unique.

At this point it must be pointed out that, in all cases where a decomposition
is possible in principle, we actually have obtained a solution for,
the system (\ref{systemofequations}). Once a
would-be solution is found, it can easily be tested by evaluating
\eqn{counttoone} for additional random values of the loop
momentum\footnote{In a certain sense, \eqn{counttoone} implies
an {\em infinite\/} number of linear equations, of which we take $\xi$ and
solve them.} This `global 1=1 test' then verifies this solution, where of course
the equality is supposed to hold only up to the precision used.

This finishes  the discussion for scalar one-loop iGraphs, 
that have unity for their numerator. Let us
regard and iGraph of order $n$ with a nontrivial numerator, for instance
\[
{q\cdot k\over D_1D_2\cdots D_N}\;\;.
\]
Now, we can always arrange for ${p_1}^\mu=0$ by a suitable
shift of the loop momentum, and write the vector $k^\mu$ as
\be
k^\mu = \omega^\mu + 2\sum_{j=2}^n \zeta_j\,{p_j}^\mu\;\;,
\ee
where the $\zeta$'s are fixed numbers and $\omega\cdot p_j=0$ for all $j$
(if $n\ge d+1$ then $\omega^\mu$ simply vanishes). We can then write
\be
(q\cdot k) = (q\cdot\omega) + \sum_{j=2}^n \zeta_j\left(
D_j-D_1-\mu_j+\mu_1\right)\;\;
\ee

so that this nonscalar iGraph decomposes into scalar iGraphs of order $n$  and
$n-1$, plus possibly a spurious term about which we do not worry since
it integrates to zero. Our treatment of the scalar case is therefore
sufficiently general.

\section{Reduction of two-loop integrands}
\subsection{Preliminaries}
We now turn to the problem of reducibility at two loops. 
Recently several attempts in this direction have appeared in the literature
(see i.e. \cite{Gluza:2010ws}, \cite{Mastrolia:2011pr}, \cite{Badger:2012dp}).
Let us assume that $l_1$ and $l_2$ are the two loop momenta. We consider three different kind 
of propagators
for the three different loop lines of a generic two loop iGraph.
\bea
D(l_1+p_i)\;,\;D(l_2+p_j)\;,\;D(l_1+l_2+p_k)
\eea
where for instance $D(l_i+p_j)=(l_i+p_j)^2-m_j^2$ and the $p_j$   are the 
external momenta associated with the propagators of the diagram.

Such iGraphs can be denoted by the triplet $(n_1,n_2,n_3)$ which indicates the
number $n_1$ of propagators that contain only the one loop momentum $l_1$,
the number $n_3$ of propagators containing only the other loop momentum
$l_2$, and the $n_2$ propagators containing both. Obviously due to the symmetries of the iGraphs, for instance exchange $l_1\leftrightarrow l_2$,
we have relations of the form
$(n_1,n_2,n_3)\leftrightarrow (n_3,n_2,n_1)$ or $(n_1,n_2,n_3) \leftrightarrow (n_1,n_3,n_2)$ provided
we also exchange properly the external momenta. Predictably,
we write the total order of the iGraph as $n=n_1+n_2+n_3$.\\
\begin{center}
\[\epsfig{figure=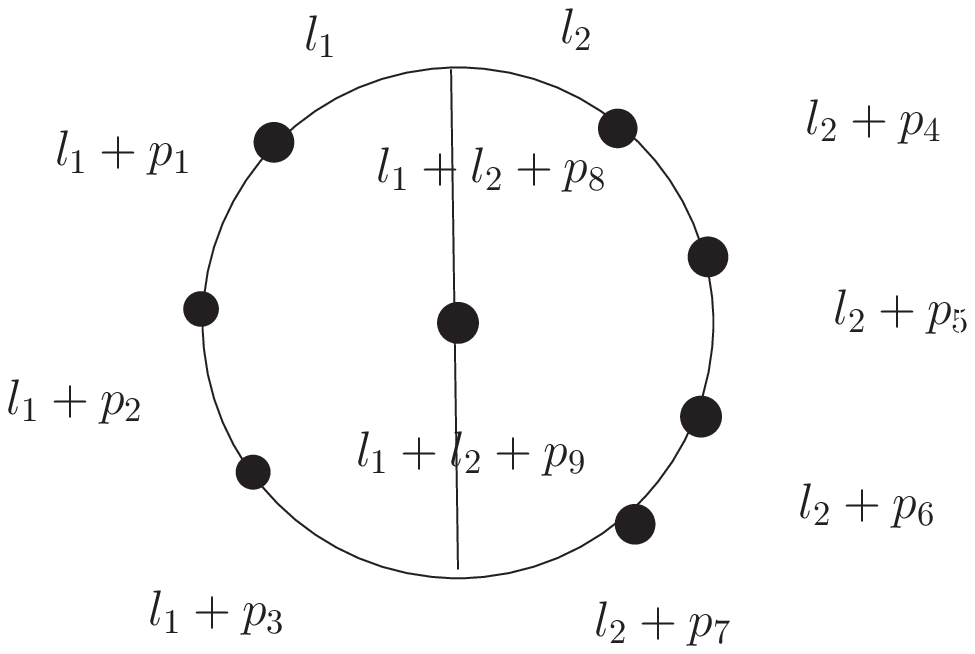,width=5cm}\]
{\small An example of the two-loop iGraph  (4,2,5).\\
One can see the three different loop lines.}
\end{center}
The propagators depending on both loop momenta are called {\em mixed propagators}. 
If these are absent the two integrals factor out and the problem 
becomes a double copy of one loop integrals.
The same happens in case any other loop line is missing since, by shifting,
one can always arrange the loop momenta such that they factor out. 
We  consider these cases solved (by the one loop techniques) and  will not discuss
them further.\\

The general strategy consists in finding function $x_j \equiv x_j(l_1,l_2)$ satisfying the following equation

\bea
\sum_{j=1}^{n_1} x_j D(l_1+p_j) +
\sum_{j=n_1+1}^{n_1+n_2} x_j  D(l_1+l_2+p_j)  + 
\sum_{j=n_1+n_2+1}^n x_j D(l_2+p_j)  = 1\;\;.
\label{counttonetwoloop}
\eea

As in the one-loop case, a generic graph of order $n\le 2d$ with $n_{1,2,3}\le 4$
cannot be decomposed in this way, since there are $l_{1,2}$  momenta for which all propagators appearing in the above equation
can be simultaneously on-shell. 
 This does {\em not\/} imply on the other hand that
 iGraphs of higher order must always be decomposable for any phase-space and mass configuration. A counterexample is
 the {\em Feynman\/} diagram of order 5 in two dimensions\footnote{This diagram
{\em can}, in fact, be decomposed, but not by the method described above:
instead one has to use IBP techniques.}: if all internal lines in this self-energy Feynman diagram are massless,
it is possible to choose the two loop momenta components such that all five propagators are simultaneously
on-shell. 
  \[\epsfig{figure=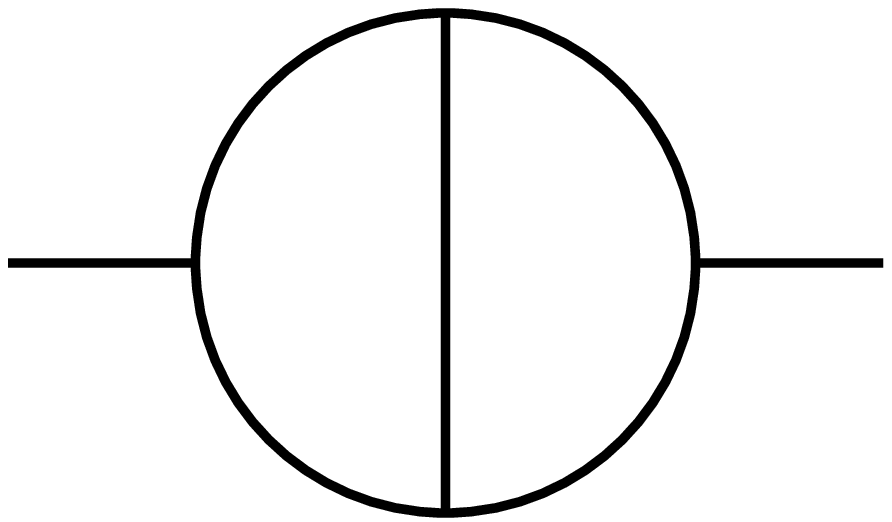,width=4cm} \]

The requirement for trivial decomposition 
(with coefficients $x_j$ that are constants with respect to the loop momenta)
now reads
\bea
\sum_{j=1}^{n_1}x_jD(l_1+p_j) +
\sum_{j=n_1+1}^{n_1+n_2}x_j D(l_1+l_2+p_j) + 
\sum_{j=n_1+n_2+1}^nx_j  D(l_2+p_j) = 1\;\;.
\label{counttonetwoloop_0}
\eea
Proceeding in analogy with our one-loop discussion, we find that we
have to satisfy the following equations :
\bea
\sum_{j=1}^{n_1+n_2}x_j = \sum_{j=n_1+1}^nx_j =
\sum_{j=n_1+1}^{n_1+n_2}x_j = 0\;\;,
\eea
\bea
\sum_{j=1}^{n_1+n_2}x_j{p_j}^\mu = 
\sum_{j=n_1+1}^{n}x_j{p_j}^\mu = 0\;\;,
\eea
and
\bea
\sum_{j=1}^nx_j\mu_j = 1\;\;.
\eea
In total there are $2d+4$ conditions, so that the minimum size of a 
trivially decomposable iGraph is $2d+4$. In four dimensions, scalar iGraphs
can therefore be decomposed down to $n=11$. Again in analogy, for $n=11$,
since by shifting we can arrange $p_1+\cdots+p_{n_1+n_2}=0$
as well as $p_{n_1+1}+\cdots+ p_n=0$, the only solution to the 
$2d+3$ homogeneous
equations is $x_j=0$, $j=1,\ldots,n$, and this fails the inhomogeneous
equation. On the other hand, since any subset of an iGraph is itself an iGraph,
any iGraph with $n_1\ge6$, $n_2\ge6$, or $n_3\ge6$ is trivially 
decomposable (for $d=4$). 
Furthermore, with linear terms we see, from the one-loop
discussion\footnote{In case there at least 6 propagators in one loop line we can 
first reduce the propagators in this loop line with constant coefficients 
and then continue further if
possible. In the case of 5 propagators we already know that 
adding linear terms that depend only on the loop
momentum of this loop line and take all coefficients that depend on the 
other loop momentum to zero,
we can again solve the problem \`a la one loop.}
that we only have to consider two-loop iGraphs with
\bea
n_{1,2,3}\le 4\;(=d)\;\;,\;\;\;n_1+n_2+n_3\le 11\;(=2d+3)\;.
\label{relevantiG}
\eea

A word of caution is in order here. We may have a case where $n_1+n_2\ge6$ or $n_2+n_3\ge6$
and then decide to perform a decomposition {\it \`a la\/} one-loop with trivial terms,
taking $l_1$ as the loop momentum, for instance, and $l_2$ as one of the "external" momenta or vice-versa. 
Since the solution of the \eqn{counttonetwoloop_0}
is nonlinear with respect to the external momenta, due to matrix
inversion, the resultant decomposition will not have, in general, the simple
form of iGraphs again, and the emerging integrals will belong to very
different classes of functions. 

The number of two-loop iGraphs that we have to consider
is therefore not very large : 4 for $d=2$, 10 for $d=3$, 19 for $d=4$.

\subsection{Further reduction with linear terms}
With trivial decomposition we see that we can always end up with an iGraph of order $2d+3$. Like in the one loop case, we now add coefficients linear in the loop 
momentum and
hope for further reductions. 

A note is in order here. 
In the one loop case the resulting integrals 
were always scalar. The reason is that any contraction of the loop 
momentum with any vector can either
reconstruct denominators or be a spurious term. 
After integrating, in the case a denominator is 
reconstructed the remaining integral is a scalar 
integral with fewer denominators. In case the term 
is spurious it vanishes after integration.
In two loops this is not the case anymore. One can always use dot products
of the loop momenta with the momenta of the integrals to write relations like
\bea
2 l_1 \cdot p_j=[(l_1+p_j)^2-m_j^2]-l_1^2-p_j^2+m_j^2\;\;.
\eea
The denominator $D(l_1+p_j)$ may, however, not be present in the integral in
case $p_j$ appears in a propagator of another type such as  $D(l_2+p_j)^2$.
Then, the product $l_1 \cdot p_j$ may be  an {\em irreducible scalar product\/} (ISP) 
\cite{Mastrolia:2011pr}. But not always; the ISP's of an integral are more complicated to write. 
For example, 
if there are enough propagators of the type $(l_1+p_i)^2$ in the diagram such 
that the $p_i$'s can form a basis,
one can rewrite $p_j$ as a linear combination of the $p_i$'s and manage to
reconstruct denominators. There
is a specific number of ISP's in any diagram and one can 
have  some freedom  in how to write them.
The integrals of the resulting basis can have numerators with 
ISP's in some power. It is not obvious that a scalar integral with a specific number
of denominators is more difficult to calculate than a non-scalar integral with fewer denominators;
however
it is commonly accepted that this is the case. 

Again, we want to write, if possible,  the number 1 as in \eqn{counttonetwoloop}. 
We use general linear terms in 
the sense that in every dimension we construct a basis $t_i$ 
(possibly, but not necessarily, the external momenta in the iGraph) and we have
\bea
x_i=a_i+\sum_jb_{ij}(l_1 \cdot t_j) +\sum_jc_{ij}(l_2 \cdot t_j)
\label{generallinear}
\eea
with the $a_i,b_{ij}$, and $c_{ij}$ constants with respect to the loop momenta.
Since in $d$ dimensions, we need $d$ vectors to construct such a basis, it is obvious that for an 
iGraph of order $n$ we start with $(2d+1)n$ coefficients. 
As in the one-loop case, we give a table
that contains, for every dimension we worked with, the number of tensor structures 
(denoted by $T(d)$) and the number of independent coefficients that we have explicitly calculated by cancellation probing as described above.
\begin{center}
$T(d) = (4d^2+18d+2)/2$\\ \[
\begin{tabular}{|c|c|c|c|c|c|c|}
\hline\hline
$n$ & $d=6$ & $d=5$ & $d=4$ & $d=3$ & $d=2$ & $d=1$ \\ \hline
3   & 39-0     & 33-0     & 27-0     & 21-0   & 15-0   & 9-0 \\ \cline{7-7}
4   & 52-0     & 44-0     & 36-0     & 28-0   & 20-0   & 12-2  \\
5   & 65-1     & 55-1     & 45-1     & 35-1   & 25-1   & 15-5 \\ \cline{6-6}
6   & 78-3     & 66-3     & 54-3     & 42-3   & 30-3   &  \\
7   & 91-6     & 77-6     & 63-6     & 49-6   & 35-8   &  \\ \cline{5-5}
8   & 104-10 & 88-10   & 72-10   & 56-10 &   &   \\
9   & 111-15 & 99-15   & 81-15   & 63-17 &   &  \\ \cline{4-4}
10  & 130-21& 110-21 & 90-21   &   &   &   \\ 
11 & 143-28 & 121-28 & 99-30   &   &   &  \\ \cline{3-3}
12 & 156-36 & 132-36 &   &   &   &   \\
13 & 169-45 & 143-47 &   &   &   &   \\ \cline{2-2}
14 & 182-55 &   &    &  &   &\\
15 & 195-55 &   &    &  &   &   \\
 \hline
$T(d)$ & 127 & 96 & 69 & 46 & 27 & 10 \\
\hline\hline
\end{tabular}\]
\end{center} 
In the table a line distinguishes between reducible 
and non-reducible cases. Reducibility is explicitly checked by \eqn{counttonetwoloop}. 
For all iGraphs defined by \eqn{relevantiG}, we see that the number of independent coefficients becomes equal to the number
of independent tensor structures, when reducibility is attained. In all dimensions every $n=2d+2 $ case is reducible with linear
terms to a
$n=2d+1$ iGraph. In four dimensions, we can decompose every integral down to integrals 
with 9 denominators. 
At this point, we are one step away from the limiting value of 8.

\subsection{Comments on the number of independent coefficients}
The most difficult part in our counting is always the number of independent coefficients. 
As shown,
we do it numerically but we would like to understand more 
the reason why we have as many 
independent coefficients as we do find. 
We try to demonstrate here a way to estimate this number; 
for the case of linear terms we will give some examples.

We can rewrite the terms in \eqn{generallinear}. As we mentioned, 
terms of the form
$l_i \cdot p_j$ either reconstruct denominators either become ISP. 
Let us assume the $(4,1,4)$ iGraph
in four dimensions. It has in principle two ISP that we call $\sigma_1$ and $\sigma_2$.
To see this, we note that in any of the loop lines that consists of four denominators, there
are three external momenta. We can always "borrow" a fourth one from the other line to have a
complete basis and write any product $l_i \cdot p_j$ as a linear combination of the four 
propagators and an ISP. The ISP in that case would be the product of $l_i$ with the 
momentum we borrowed. Repeating for the other loop line we get the two ISP's.
Using the coefficients of \eqn{generallinear}
in \eqn {counttonetwoloop} we can either get a denominator times a 
constant or an ISP, or products 
of two denominators. 
We write this equation schematically as
\bea
1=\sum_{i=1}^{9} D_i\{1,\sigma_1,\sigma_2\} +\sum_{i=1}^9 \sum^9_{j,j \geq i}D_iD_j \{1\}
\label{alternativecounting}
\eea
This  means that by writing $\{1,\sigma_1,\sigma_2\}D_i$  
there is a constant coefficient in front of every of these different terms:
\[\{1,\sigma_1,\sigma_2\}=a\cdot 1+b_1\sigma_1 +b_2 \sigma_2\]
where a, $b_1$, $b_2$ general numbers.
In our particular example,
we have $9\times3 +45\times 1=72$ coefficients. However, we started with a problem with up to
cubic power in loop momenta and ended with up to quartic powers since we have these products 
of two denominators times some constants. 
These higher powers have to cancel, which means that we have to
put extra constraints on our coefficients. We have 6 such constraints to cancel, namely the
\[l_1^4,\,\, l_1^2l_2^2, \,\,l_2^4,\,\,(l_1 \cdot l_2)l_1^2,\,\,
 (l_1 \cdot l_2)l_2^2,\,\,(l_1 \cdot l_2)^2\] 
terms. As a result we end up with $72-6=66$ independent coefficients, which is the number we get 
numerically as well.
If we now try to decompose an iGraph of order 10 we can prove that 
\eqn{alternativecounting} becomes 
\bea
1=\sum_{i=1}^{10} D_i(1,\sigma_1,\sigma_2) +\sum_{i=1}^9 \sum^9_{j,j \geq i}D_iD_j (1)
\eea
We don't need to go up to 10 in the product of 2 denominators since \footnote{This 
is actually the point
where the number of dimensions plays a role in the counting.}
\[D_{10} \propto (D_1,...,D_9,\sigma_1,\sigma_2) \]
We still need them, though, in the first term to produce terms of 
the type $(\sigma_i)^2$. In that case 
we have 69 independent coefficients and this graph is reducible. 
Adding more propagators we do not get more  
independent coefficients. In the same way one can count
the independent coefficients in all dimensions although it is clear that it is safer 
to find their number numerically since there are a lot of overlaps in the 
tensor structures for higher cases. The way of rewriting the general linear terms 
as propagators and ISP's  in the example above is still not the OPP method. In an extension
of the OPP method to two loops, one would find the ISP's of every subdiagram and would avoid terms like
$D_i^2$. We expect something similar to the one-loop case to happen then, rewriting 
$\sigma_1$ and $\sigma_2$ in the form of true ISP's  of every subdiagram's contributions
of the terms with the highest number of denominators to cancel. For that it is possible
that special properties exist, as again in the one-loop case where spurious term solve
 a lot of equations by putting automatically tensor structures to zero.

\subsection{Reduction with general quadratic terms}
We try to reduce our iGraphs further with the use of general quadratic terms in the coefficients. 
In this case the coefficients become
\bea
&&x_i=a_i+\sum_jb_{ij}(l_1 \cdot t_j) +\sum_jc_{ij}(l_2 \cdot t_j) +
\sum_{j\le k}d_{ijk}
(l_1 \cdot t_j)(l_1 \cdot t_k) \,\nn
&& +\sum_{j\le k}e_{ijk}(l_2 \cdot t_j)(l_2 \cdot t_k) +
\sum_{j,k}f_{ijk}(l_1 \cdot t_j)(l_2 \cdot t_k)) \, \nn
\label{generalquadratic}
\eea
We give 
the number of tensor structures $T(d)$ and  the original number of coefficients with general quadratic terms, $C_1(d)$.  The coefficients 
depend on the number of propagators $n$. Notice that the expression for
 $T(d)$ is not valid for $d=2$. In that case, there is more overlap between the
 highest tensor structures. More specifically, for this particular case one can completely 
reconstruct the $l_1^2l_2^{\mu}l_2^{\nu}$
structure from $l_2^2l_1^{\mu}l_1^{\nu}$ and $(l_1 \cdot l_2)l_1^{\mu}l_2^{\nu}$ allowing fewer independent 
structures to be constructed.

\bea T(d) &=& {4d^3/3}+10d^2+{20d}/{3}-2  \\
C_1(d) &=&(2d^2+3d+1)n 
\eea

With quadratic terms we start with $(2d^2+3d+1)\times n$ coefficients in total, 
not all of them 
of course being independent. Using cancellation probing we are able to find the number of 
independent coefficients for different iGraphs in different dimensions. We put our findings in
the following table :
\begin{center}
\[
\begin{tabular}{|c|c|c|c|}
\hline\hline
$n$ & $d=4$    & $d=3$     & $d=2$  \\ \hline
3     & 135-4     & 84-3       & 45-3   \\ 
4     & 180-6     & 128-6     & 60-6   \\  \cline{4-4}
5     & 225-18   & 140-16   & 75-15   \\ 
6     & 270-38   & 168-32   & 90-30 \\ 
7     & 315-65   & 196-53   &             \\ \cline{3-3}
8     & 360-98   & 224-80   &          \\ 
9     & 405-136 & 252-108 &           \\ \cline{2-2}
10   & 450-180 &               &           \\ 
11   & 495-225 &               &            \\ 
 \hline
$T(d)$ & 270 & 144 & 60  \\
\hline\hline
\end{tabular}\]
\end{center} 

In the table above a line distinguishes again between reducible 
and non-reducible cases, according to \eqn{counttonetwoloop}.
As we see, we can decompose in 2 dimensions an iGraph of order 
5 to lower order iGraphs as indicated by unitarity.
However,
there is no such solution in 3 and 4 dimensions and in this case we have 
to investigate what happens if we 
add cubic, quartic terms and so on. The two-dimensional case is exceptional
here because of the extra properties that lower the number of tensor structures.

\subsection {Reduction with general cubic terms}
We focus now on 3 and 4 dimensions since we finished the reduction in 2 dimensions. 
We include
general cubic terms and our coefficients become
\bea
&&x_i=a_i+\sum_jb_{ij}(l_1 \cdot t_j) +\sum_jc_{ij}(l_2 \cdot t_j) +
\sum_{j\le k}d_{ijk}
(l_1 \cdot t_j)(l_1\cdot t_k)\nn &&  +
\sum_{j\le k}e_{ijk}(l_2 \cdot t_j)(l_2\cdot t_k) 
+\sum_{j,k}f_{ijk}(l_1 \cdot t_j)(l_2 \cdot t_k)\nn && + 
\sum_{j\le k\le l}g_{ijkl}(l_1 \cdot t_j)(l_1\cdot t_k)(l_1\cdot t_l) 
+\sum_{l}\sum_{j\le k}h_{ijkl}(l_1 \cdot t_j)(l_1\cdot t_k) (l_2 \cdot t_l) 
\nn &&+ \sum_l\sum_{j\le k}p_{ijkl}(l_2 \cdot t_j)(l_2\cdot t_k) (l_1 \cdot t_l)
 + 
\sum_{j\le k\le l}q_{ijkl}(l_2 \cdot t_j)(l_2\cdot t_k)(l_2\cdot t_l)) \, \nn
\label{generalcubic}
\eea
We give 
the number of tensor structures $T(d)$ and  the original number of coefficients 
with general cubic terms
$C_1(d)$. 
The coefficients 
depend on the number of propagators $n$. 
The expression for $T(d)$ is valid for $d \geq3$.
 
\bea T(d) &=& {2d^4}/{3}+{22d^3}/{3}+{71d^2}/{6}+{d}/{6}+1  \\ 
C_1(d) &=& ({4d^3}/{3}+4d^2 + {11d}/{3} +1)n \eea

This means that in $d$ dimensions we start with $C_1(d)$
coefficients, not all of them being independent.
We run the MAPLE code in the case of the iGraph of order 7 in 3 dimensions and we find that
out of 588
original coefficients, 360  are independent. This is the number of tensor 
structures as well. Using another PYTHON-based program\footnote{http://codepad.org/zT4wUxCJ}, we can
actually solve the system decomposing any iGraph of order 7 in 3 dimensions to lower iGraphs
with general cubic terms, and perform the 1=1 test. This
means that with cubic terms we are able to 
decompose any two-loop
iGraph in 3 dimensions to up to a $2d$ iGraph as expected from unitarity. 
In the same way, we can investigate $d=4$, and we get a valid decomposition:
from our original
1485 coefficients, 831 are independent and all the tensor structures can be reconstructed.
Actually we did the same in 5, 6, 7, and 8
 dimensions, and managed to decompose every 
integral of generic order $2d+1$ or higher, using cubic terms. 
We believe that this is a general result
for any dimension, except of course for $d=2$, where the same can be achieved with only quadratic terms.

\section{ Conclusions}
By introducing the notion of iGraphs,
we have investigated the decomposability of 
Feynman amplitudes at one and two loops.
In both cases we have demonstrated (in the two-loop case, up to $d=8$) that
generic iGraphs can be decomposed down to the unitarity-based limit
of $n=d$ and $n=2d$, respectively. For one-loop graphs, the inclusion of
linear terms is sufficient, and we have elucidated the relation between
these linear terms and the OPP's spurious terms. At the two-loop level,
ultimately cubic terms are needed (quadratic for $d=2$), and the decomposition
is seen to lead to non-scalar, non-vanishing integrals. If one wants to design a
two-loop OPP method, it is clear that one has 
 to take our general linear, quadratic, cubic terms and rewrite them in terms of propagators 
 (this would lead to contributions with less denominators) and ISP's of each subdiagram 
 seperately
 \footnote{ In the same way that one-loop OPP has different spurious terms for every 
 subdiagram}. Our work is basically the starting point of an OPP method and a proof
 that reductions that were conjectured in 
 \cite{Mastrolia:2011pr,Badger:2012dp} are actually valid and survive the global
 $1=1$ test, in case one includes all relevant cuts.

We note once again that the resulting integral basis, obtained in this way, is clearly not a minimal one.
We are aware of cases that could be further decomposed using IBP
identities. Achieving a proper level of understanding of the interplay between OPP and IBP will certainly open
the road for an efficient reduction of two-loop amplitudes at the integrand level. 
\\

\noindent {\bf Acknowledgement:} We acknowledge useful discussions with Prof. M. Czakon.

\appendix
\section{ Appendix: Pentagons to Boxes with the OPP method}
By defining

\bea
D_n(p_1,p_2,...,p_n)=\frac{1}{D(q+p_1) D(q+p_2)...D(q+p_n)}
\eea
we try to decompose this pentagon

\[ D_5(p_1,p_2,p_3,p_4,p_5) \]
 to the 5 following boxes

\[D_4(p_2,p_3,p_4,p_5),D_4(p_3,p_4,p_5,p_1),D_4(p_4,p_5,p_1,p_2) \]
\[D_4(p_5,p_1,p_2,p_3)D_4(p_1,p_2,p_3,p_4) \]
By shifting the loop momenta we can always arrange that the momenta $p_i$ of the denominators 
sum up to zero:
\[ \sum_{i=1}^5 p_i=0 \]
We write the spurious terms in the following form

\bea
&&S_1=\epsilon(q+p_2,q+p_3,q+p_4,q+p_5)=q_{\mu}(\epsilon^{\mu}(p_3,p_4,p_5)- 
\epsilon^{\mu}(p_2,p_4,p_5) + \nl
&&\epsilon^{\mu}(p_2,p_3,p_5)-\epsilon^{\mu}(p_2,p_3,p_4))+ \epsilon(p_2,p_3,p_4,p_5) = \nl
&&q_{\mu}A_1^{\mu} +B_1 \nl
\eea
We then have:

\bea
&&A_1^{\mu}=\epsilon^{\mu}(p_3,p_4,p_5)-\epsilon^{\mu}(p_2,p_4,p_5)+
\epsilon^{\mu}(p_2,p_3,p_5)-\epsilon^{\mu}(p_2,p_3,p_4) \nonumber \\
&&A_2^{\mu}=\epsilon^{\mu}(p_4,p_5,p_1)-\epsilon^{\mu}(p_3,p_5,p_1)+
\epsilon^{\mu}(p_3,p_4,p_1)-\epsilon^{\mu}(p_3,p_4,p_5) \nonumber \\
&&A_3^{\mu}=\epsilon^{\mu}(p_5,p_1,p_2)-\epsilon^{\mu}(p_4,p_1,p_2)+
\epsilon^{\mu}(p_4,p_5,p_2)-\epsilon^{\mu}(p_4,p_5,p_1) \nonumber \\
&&A_4^{\mu}=\epsilon^{\mu}(p_1,p_2,p_3)-\epsilon^{\mu}(p_5,p_2,p_3)+
\epsilon^{\mu}(p_5,p_1,p_3)-\epsilon^{\mu}(p_5,p_1,p_2) \nonumber \\
&&A_5^{\mu}=\epsilon^{\mu}(p_2,p_3,p_4)-\epsilon^{\mu}(p_1,p_3,p_4)+
\epsilon^{\mu}(p_1,p_2,p_4)-\epsilon^{\mu}(p_1,p_2,p_3) \nonumber \\
\label{as}
\eea
Using the OPP master formula  \cite{Ossola:2006us}  we get 

\bea
&&1=[d_1+\tilde d_1 S_1]D(q+p_1) +[d_2  +\tilde d_2S_2]D(q+p_2)+...+ \nl
&&[d_5 +\tilde d_5S_5]D(q+p_5) \nl
\label{penta}
\eea

We compare the polynomials of the two sides of \eqn{penta}. The $q^2 q_{\mu}$ terms must vanish
for any value of $q$ which means

\bea
\sum_{i} (\tilde d_i A_i^{\mu})=0
\label{highest}
\eea

Notice that from \eqn{as}

\[\sum_{i=1}^5 A_i =0 \] 

which means that if  all $\tilde d_i$ are equal 
\eqn{highest} is satisfied. 

This is actually the only solution. To see that, consider the sum 
\[\tilde d_1 A_1^{\mu}+ \cdots +\tilde d_5 A_5^{\mu} \] 
Substituting 

\[p_5=-p_1-p_2-p_3-p_4 \] we get

\bea
\sum(\tilde d_i A_i^{\mu})&=&(\tilde d_1+\tilde d_2-4 \tilde d_3+\tilde d_4+\tilde d_5) \epsilon^{\mu}(p_1,p_2,p_4)  \nonumber \\
&+& (-\tilde d_1+4\tilde d_2-\tilde d_3-\tilde d_4-\tilde d_5)\epsilon^{\mu}(p_1,p_3,p_4) \nonumber \\
&+&(-\tilde d_1-\tilde d_2-\tilde d_3+4\tilde d_4-\tilde d_5)\epsilon^{\mu}(p_1,p_2,p_3)  \nonumber \\
&+&(-4\tilde d_1+\tilde d_2+\tilde d_3+\tilde d_4+\tilde d_5)\epsilon^{\mu}(p_2,p_3,p_4)  \nonumber \\
\eea

In 4 dimensions, for this sum to be zero, all coefficients in front of the vectors are zero (the vectors are 
linearly independent). This results in the system:

\bea
&&\tilde d_1+ \tilde d_2-4\tilde d_3+\tilde d_4+\tilde d_5 =0 \nonumber \\
&&-\tilde d_1+4\tilde d_2-\tilde d_3-\tilde d_4-\tilde d_5 =0 \nonumber \\
&&-\tilde d_1-\tilde d_2-\tilde d_3+4\tilde d_4-\tilde d_5=0  \nonumber \\
&&-4\tilde d_1+\tilde d_2+\tilde d_3+\tilde d_4+\tilde d_5 =0 \nonumber \\
\eea

One can solve the system and find 
\bea
\tilde d_1=\tilde d_2=\tilde d_3=\tilde d_4= \tilde d_5 
\label{dtildes}
\eea

The only solution is when all coefficients are equal and 
from now on we call them $\tilde d$. 
Then we take a 
look at the quadratic in $q$ parts
of the right-hand side of \eqn{penta}. They cancel as well but one has to be careful since they
come from two terms, the $q^2$ and the $q^{\mu}q^{\nu}$ term. We look at the latter term

\[ \tilde d q_{\mu} \sum_{i=1}^5 (q\cdot p_i)A_i^{\mu} \]

We have :

\bea
&&\tilde d q_{\mu} \sum_{i=1}^5 (q\cdot p_i)A_i^{\mu}= \tilde d q_{\mu} ( (A_1-A_5)(q \cdot p_1)+\nonumber \\
&&(A_2-A_5) (q \cdot p_2)+(A_3-A_5) (q \cdot p_3)+(A_4-A_5) (q \cdot p_4) ) \nonumber \\
\eea

where $p_5=-p_1-p_2-p_3-p_4$ is again used. With this substitution we have:

\bea
&&A_1-A_5=-5 \epsilon^{\mu}(p_2,p_3,p_4) \nonumber \\
&&A_2-A_5=5 \epsilon^{\mu}(p_1,p_3,p_4) \nonumber \\
&&A_3-A_5=-5 \epsilon^{\mu}(p_1,p_2,p_4) \nonumber \\
&&A_4-A_5=5 \epsilon^{\mu}(p_1,p_2,p_3) \nonumber \\
\eea

Inserting the Schouten Identity 

\bea
&&\epsilon(p_1,p_2,p_3,p_4)q^{\mu}= \epsilon^{\mu}(p_2,p_3,p_4)(q \cdot p_1) -
\epsilon^{\mu}(p_1,p_3,p_4)(q \cdot p_2) + \nl
&&\epsilon^{\mu}(p_1,p_2,p_4)(q \cdot p_3)- \epsilon^{\mu}(p_1,p_2,p_3)(q \cdot p_4) \nl
\eea

 we get 

\bea
\tilde d q_{\mu} \sum_{i=1}^5 (q\cdot p_i)A_i^{\mu}=-5\tilde d q^2 \epsilon (p_1,p_2,p_3,p_4)
\eea

That is exactly the property that the spurious terms have that makes the solution to the system
possible. The $q^{\mu}q^{\nu}$ terms all vanish owing to the Schouten identity except
for the trace part proportional to $q^2$  solving 9 out of 10 equations in one go. The total number
of nontrivial equations is therefore 10 and not 19 in this case and the system has a solution.
To complete the story, taking into account \eqn{dtildes} we are left with 5 $d_i$ and one $\tilde d$, which are uniquely now determined
from the remaining 6 equations, namely constant part, $q^2$ term and $q_{\mu}$ term.


\providecommand{\href}[2]{#2}\begingroup\raggedright\endgroup

\end{document}